\documentclass[conference]{IEEEtran}
\IEEEoverridecommandlockouts

\usepackage{cite}
\usepackage{amsmath,amssymb,amsfonts}
\usepackage{algorithmic}
\usepackage{graphicx}
\usepackage{textcomp}
\usepackage{xcolor}

\usepackage{cuted}
\usepackage{graphicx}
\usepackage{booktabs}
\usepackage{amsmath}
\usepackage{amssymb}
\usepackage{multirow}
\usepackage{arydshln}
\usepackage{xcolor}
\usepackage{pifont}
\usepackage{wrapfig}
\usepackage{array} 
\usepackage[breaklinks,colorlinks,citecolor=cyan,linkcolor=blue]{hyperref}
\usepackage{cleveref}
\usepackage{enumitem}

\def\BibTeX{{\rm B\kern-.05em{\sc i\kern-.025em b}\kern-.08em
    T\kern-.1667em\lower.7ex\hbox{E}\kern-.125emX}}
\begin{document}



\title{Sequential Contrastive Audio-Visual Learning}

\author{\IEEEauthorblockN{Ioannis Tsiamas$^{\dagger *}$\thanks{$\dagger$ Work done during an internship at Dolby Laboratories.}\thanks{$*$ Corresponding author at \footnotesize\texttt{ioannis.tsiamas@upc.edu}}}
\IEEEauthorblockA{\textit{Universitat Politècnica de Catalunya}}
\and
\IEEEauthorblockN{Santiago Pascual, Chunghsin Yeh, Joan Serrà}
\IEEEauthorblockA{\textit{Dolby Laboratories}}
}

\maketitle

\begin{abstract}
    Contrastive learning has emerged as a powerful technique in audio-visual representation learning, leveraging the natural co-occurrence of audio and visual modalities in web-scale video datasets. However, conventional contrastive audio-visual learning (CAV) methodologies often rely on aggregated representations derived through temporal aggregation, neglecting the intrinsic sequential nature of the data. This oversight raises concerns regarding the ability of standard approaches to capture and utilize fine-grained information within sequences. In response to this limitation, we propose sequential contrastive audio-visual learning (SCAV), which contrasts examples based on their non-aggregated representation space using multidimensional sequential distances. Audio-visual retrieval experiments with the VGGSound and Music datasets demonstrate the effectiveness of SCAV, with up to 3.5$\times$ relative improvements in recall against traditional aggregation-based contrastive learning and other previously proposed methods, which utilize more parameters and data. We also show that models trained with SCAV exhibit a significant degree of flexibility regarding the metric employed for retrieval, allowing us to use a hybrid retrieval approach that is both effective and efficient.
\end{abstract}

\begin{IEEEkeywords}
Audio-Visual, Contrastive, Multimodality.
\end{IEEEkeywords}

\section{Introduction} \label{sec:intro}

\begin{figure}[t]
        \centering
        \includegraphics[width=0.8\columnwidth]{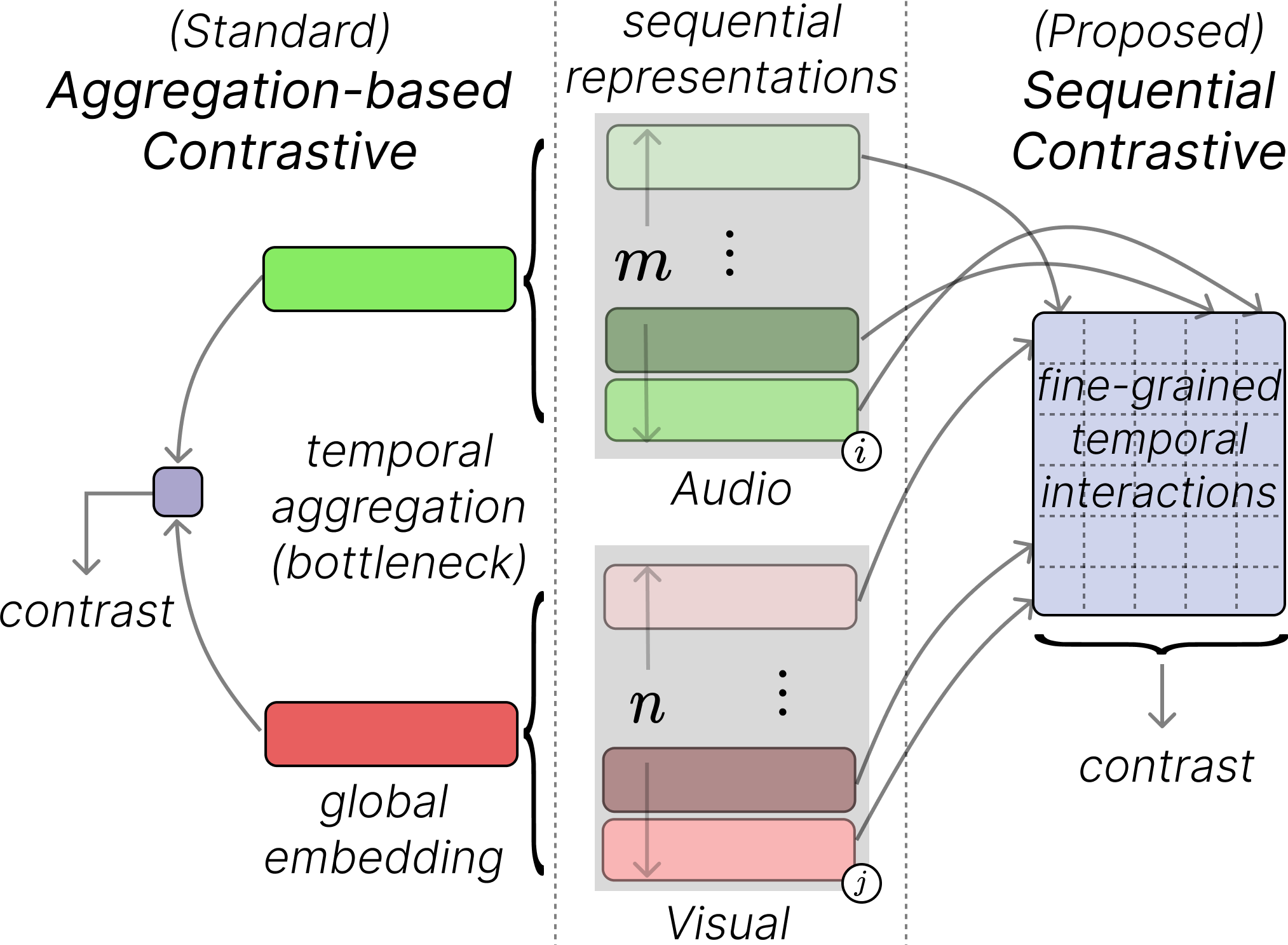}
        \caption{Aggregation- vs.\ sequence-based contrastive learning.}
        \label{fig:intro}
        \vspace{-0.2cm}
    \end{figure}

    Audio-visual representation learning plays a central role in several recent advancements such as multimodal LLMs~\cite{gemini,llama3} and audio-visual generative models~\cite{power_of_sound,diff-foley,maskvat,any-to-any}. Contrastive learning~\cite{contrastive_loss,infonce} has emerged as an effective methodology for learning audio-visual representations by relying on the co-occurrence of the two modalities in unlabeled web-scale video datasets. The conventional approach involves contrasting between (aggregated) global embedding vectors~\cite{cooperative_learning,cav_mae}, which should ideally capture the semantics of each example (Fig.~\ref{fig:intro}, left). Although such a global embedding may be sufficient for static modalities like images~\cite{clip}, we argue that it is potentially over-compressing for modalities of dynamic nature like videos or music, thus hindering the effectiveness of contrastive learning and the robustness of the multimodal space. For example, such a model would struggle in an audio-visual retrieval setting with music videos. Key information, such as a riff's tempo or the hand movements of a guitarist, would be mostly lost due to the compression into a single non-temporal embedding.

    In this paper, we propose sequential contrastive audio-visual learning (SCAV), which utilizes sequential distances to contrast directly on the natural and non-aggregated representation space, taking advantage of the fine-grained temporal and semantic information of the sequences (Fig.~\ref{fig:intro}, right). We experiment with different sequential distances and find that a Euclidean-based distance with interpolation is more effective and scales better to larger batch sizes than other more complex alternatives like DTW~\cite{dtw1} or Wasserstein~\cite{wasserstein}. SCAV can further utilize the sequential distance objective for retrieval, where in VGGSound~\cite{vggsound} it surpasses standard contrastive methods and other state-of-the-art models from the literature~\cite{cav_mae,imagebind} by large margins, while relying on fewer training parameters and less data. Our proposed method also excels in a challenging zero-shot retrieval setting using the Music dataset~\cite{music1}, for which strong capabilities of intra-sequence modeling and discrimination are required, due to the presence of many examples with very similar overall semantics. Finally, we propose a hybrid retrieval approach that combines the best of both the standard aggregation-based and sequence-based retrieval methods, thus providing a flexible trade-off between accuracy and efficiency.

\section{Relevant Research} \label{sec:relevant_works}

    Contrastive learning has shown significant improvements in multimodal representation learning using large-scale, text-labeled weakly supervised data, with several works on vision-language~\cite{clip,align}, audio-language~\cite{laion_clap,ms_clap}, and with extensions to more than two modalities~\cite{imagebind,audioclip}. In audio-visual representation learning, early works leveraged the natural audio-visual correspondence in unlabeled clips with objectives that predict whether an audio and a video frame belong to the same clip~\cite{look_listen_learn,objects_that_sound}, or whether they are synchronized~\cite{cooperative_learning,audio_visual_scene,audio_visual_objects_from_sound}. Later, a fully-contrastive framework was adopted in \cite{cma}, with subsequent improvements in handling negative examples~\cite{active_contrastive,on_negative_sampling} and addressing false positives and false negatives~\cite{robust_audio_visual,negative_aware}. More recently, \cite{cav_mae,mavil} demonstrated 
    considerable improvements by coupling contrastive learning with masked data modeling. Also, \cite{diff-foley,contrastive_temp} aim to enhance the temporal-awareness of the representations by combining the standard contrastive objective with additional temporal ones, focusing on speed, order, direction and synchronicity. Our work has the same goals, but we accomplish them by proposing a sequence-based contrastive objective that captures both the global semantics and the fine-grained temporal information, while it also allows us to use it for retrieval, achieving large performance gains. Although we are the first to apply sequential distances within a contrastive framework for audio-visual representation learning, there are works that have employed them for video-text pretraining~\cite{temp_clr}, visual place recognition~\cite{vpr}, action representation learning~\cite{seqcl_action}, and text recognition~\cite{seqcl_text}.
    

\section{Methodology} \label{sec:meth}

    \subsection{Model Architecture} \label{sec:model_arch}

\begin{figure}[t]
           \centering
           \includegraphics[width=0.85\columnwidth]{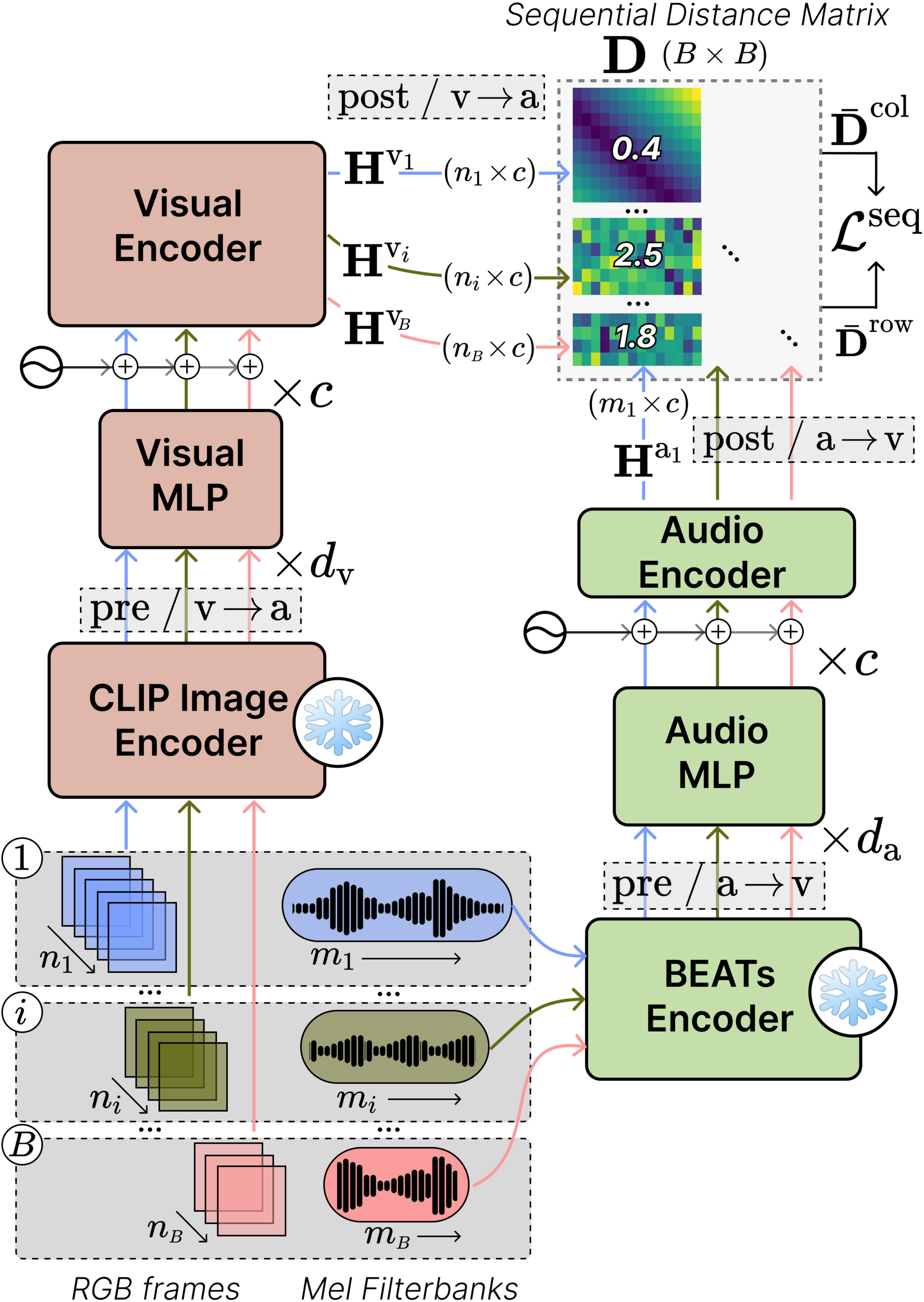} 
           \caption{Model Architecture for SCAV.}
           \label{fig:architecture}
           \vspace{-0.2cm}
        \end{figure}

        Given an unlabeled video $i$, we extract visual and audio sequences $\textbf{V}^i \! \in \! \mathbb{R}^{n_i \times h \times w \times 3}$ ($n_i$ RGB frames) and $\textbf{A}^i \! \in \! \mathbb{R}^{m_i \times b}$ ($m_i$ Mel frames). We then obtain features $\mathbf{\hat{V}}^i \! \in \! \mathbb{R}^{n_i \times d_\text{v}}$ and $\mathbf{\hat{A}}^i \! \in \! \mathbb{R}^{m_i \times d_\text{a}}$ via (frozen) pretrained models CLIP~\cite{clip} and BEATs~\cite{beats}, where generally $n_i \! \neq \! m_i$. Both feature sequences are projected to a common dimensionality $c$ using MLPs, and after adding relative positional embeddings~\cite{electra}, two Transformer encoders~\cite{transformer} (of different complexity) are used to obtain the sequential latent representations $\textbf{H}^{\text{v}_i} \! \in \! \mathbb{R}^{n_i \times c}$ and $\textbf{H}^{\text{a}_i} \! \in \! \mathbb{R}^{m_i \times c}$ (Fig.~\ref{fig:architecture}). The visual encoder is necessary because the output $\textbf{V}^i$ of CLIP is a set of independent frame representations instead of a well-formed sequence.\footnote{CLIP encodes each frame independently from the rest.} On the other hand, the output $\textbf{A}^i$ of BEATs is a proper sequence representation, but we found some additional processing to be beneficial due to BEATs being frozen and trained on a different task (audio tagging).

    \subsection{Aggregation-based Contrastive Learning} \label{sec:vanilla_contrastive}
        
        Using the latent representations obtained with the model described in \S\ref{sec:model_arch}, the standard aggregation-based contrastive loss~\cite{cav_mae,imagebind,contrastive_temp} with $B$ examples is be defined as\small
        \begin{align} \label{eq:vanilla_loss}
            \mathcal{L}^\text{agg} \!=\! - \frac{1}{2B} \! \sum_{i=1}^{B} \! \Biggl[\! &\log \! \frac{ \exp (s_{ii} / \tau) }{\sum_{j=1}^{B} \! \exp (s_{ij}/ \tau) } \! + \! \log \! \frac{ \exp (s_{ii} / \tau) }{\sum_{j=1}^{B} \! \exp (s_{ji}/ \tau) } \! \Biggr] , \nonumber
        \end{align} \normalsize
        where $\tau \! > \! 0$ is a learnable temperature parameter and $s_{ij}$ is the cosine similarity $\texttt{cos}\!: \! \mathbb{R}^c \times \mathbb{R}^c \! \rightarrow \! \mathbb{R}$ between $\mathbf{\bar{h}}^{\text{v}_i}$ and $\mathbf{\bar{h}}^{\text{a}_j}$, which are the global $i$-th video and $j$-th audio embeddings. These embeddings are derived via temporal aggregation~(i.e. mean pooling) from $\textbf{H}^{\text{v}_i}$ and $\textbf{H}^{\text{a}_i}$, respectively. 
        
        Although this approach can match the general semantics of the audio-visual scenes, it neglects the sequential nature of the visual and audio modalities due to the temporal aggregation. This is potentially limiting the capacity of the representation space to encode fine-grained temporal information, which is critical in disambiguating between semantically similar but temporally different audio-visual scenes.

    \subsection{Sequential Contrastive Learning} \label{sec:seq_contrastive}

        Due to the issues arising from collapsing the sequential representations to global embeddings (\S\ref{sec:vanilla_contrastive}), we propose a contrastive approach based on a distance computed on the non-aggregated sequential latent representation space, as in:\small
        \begin{align}
            \mathcal{L}^{\text{seq}} \! = \! - \frac{1}{2B} \! \sum_{i=1}^{B} \! \Biggl[\! \log \! \frac{ \exp (\!-\bar{d}^{\text{v}\rightarrow\text{a}}_{ii} \! / \! \lambda) }{\sum_{j=1}^{B} \! \exp(\!-\bar{d}^{\text{v}\rightarrow\text{a}}_{ij} \! / \! \lambda) } \! + \! \log \! \frac{ \exp (\!-\bar{d}^{\text{a}\rightarrow\text{v}}_{ii} \! / \! \lambda) }{\sum_{j=1}^{B} \! \exp (\!-\bar{d}^{\text{a}\rightarrow\text{v}}_{ji} \! / \! \lambda) } \! \Biggr]\!, \nonumber
        \end{align}\normalsize
        where $\lambda \! > \! 0$ is a learnable temperature and $d_{ij} \! \geq \! 0$ is the result of a sequential distance function $\texttt{dist} \!: \! \mathbb{R}^{n_i \times c} \times \mathbb{R}^{m_j \times c} \!
        \rightarrow \! \mathbb{R}^+$ between $\textbf{H}^{\text{v}_i}$ and $ \textbf{H}^{\text{a}_j}$, with $n_i \! \neq \! m_j$. After computing the $B \! \times \! B$ pairwise distance matrix $\textbf{D}$ in a batch, we use $\bar{\textbf{D}}^{\text{v}\rightarrow\text{a}}$ and $\bar{\textbf{D}}^{\text{a}\rightarrow\text{v}}$ in $\mathcal{L}^{\text{seq}}$, which are row- and column-wise z-score normalized versions of $\textbf{D}$. Although this makes the distance matrix non-symmetric, we observed this to be crucial, since it magnifies the distances between correct and in-correct pairs, which tend to diminish due to the very high-dimensional space~$(n_i \! \times \! c) \! \times \! (m_j \! \times \! c)$ 
        in which they are computed.
        
        For $\texttt{dist}$ we consider three different distance functions: (1) \emph{Interpolated Euclidean}, where we interpolate one of the sequences to the length of the other, and calculate the average squared euclidean distance between all paired points in the audio-visual representations. We experiment with different variants regarding the direction of the interpolation $(\text{v} \! \rightarrow \! \text{a}$ or $\text{a} \! \rightarrow \! \text{v}$) and where it is applied (pre or post), as indicated with gray boxes in Fig.~\ref{fig:architecture}. (2): \emph{Dynamic Time Warping} (DTW)~\cite{dtw1}, which is the distance associated with the shortest alignment path between the two sequences, where we use its differentiable soft variant~\cite{soft_dtw} during training. (3) \emph{Wasserstein}~\cite{wasserstein,ot}, which is the minimum transportation cost over all possible transportation plans between the sequences, where we follow \cite{zeroswot} to make it differentiable and permutation sensitive.

    \subsection{Retrieval with Sequential Distances} \label{sec:retrieval}

    To use a contrastively trained model for audio-visual retrieval in a set of $K$ examples, one has to apply temporal mean-aggregation to obtain $c$-dimensional embeddings, and retrieve the most similar audio/video based on cosine similarity (\emph{Aggregation-based} retrieval). But since SCAV models have been trained to minimize sequential distances between correct audio-visual pairs, we propose to also apply the same objective for retrieval, which we call \emph{Sequence-based} retrieval. Although we expect it to be more accurate, it comes with a computational overhead due to searching in a multidimensional sequential space, rather than the collapsed space of aggregation-based retrieval. To ease this overhead, we also propose a \emph{Hybrid} variant which combines the best of both methods. Specifically, we initially perform a pre-selection of $k$ candidates~(with $k \! < \! K$) using the efficient aggregation-based approach, and then obtain the best candidate using sequence-based retrieval on the filtered pool of $k$ candidates.

\section{Experimental Setup} \label{sec:experimental_setup}

    \setlength{\parindent}{0pt}

    \textbf{Data.} We use VGGSound~\cite{vggsound} to train and test our models, which contains 10-second clips from YouTube. We use a subset of the train and test splits, with 153\,k and 13\,k clips each.\footnote{This is due to missing or low-quality videos.} For validation we sample 550~videos from the train split. For out-of-distribution testing we use the Music dataset~\cite{music1,music2}, which requires strong audio onset detection from the close-up camera recording of people playing musical instruments. We extracted 1.9\,k non-overlapping clips from the 103 videos of the MUSIC21-solo test partition.\footnote{\href{https://github.com/roudimit/MUSIC_dataset}{github.com/roudimit/MUSIC\_dataset}}
    
    \textbf{Model Architecture.} We extract the RGB visual frames of each video at its native sampling rate and use CLIP~\cite{clip} to process each frame independently. We used the ViT-B/32 CLIP image encoder which is a Vision-Transformer (ViT)~\cite{vit}, with 12 blocks of dimensionality $d_\text{v}$ of 768.\footnote{\href{https://huggingface.co/openai/clip-vit-base-patch32}{huggingface.co/openai/clip-vit-base-patch32}}. For the audio we resample to 16\,kHz and extract 128-dimensional Mel Filterbank features with a window of 25\,ms, which are processed by BEATs~\cite{beats}. We used the iter3+~(AS2M)~(cpt2) model, which is a 12-layer ViT with dimensionality of 768..\footnote{\href{https://github.com/microsoft/unilm/tree/master/beats}{github.com/microsoft/unilm/tree/master/beats}} We discard the audio classification head, and use the final hidden representation, which we reconstruct to a dimensionality $d_\text{a}$ of 6144 by stacking across the patch dimension of size 8. Most visual streams from the VGGSound clips have higher sampling rates than the standard 6.2\,fps of BEATs, resulting into video sequences longer than the audio ones ($n_i \! > \! m_i$). Both CLIP and BEATs remain frozen in all our experiments. The visual and audio MLPs have hidden dimensions of 2048 and 768, respectively, and project to a common dimensionality $c$ of 512. We tuned the size of the visual and audio Transformers to 8 and 2 blocks, respectively, based on the validation set of VGGSound. They have a common dimensionality $c$ of 512, making a total number of training parameters of 38\,M. 
    
    \textbf{Training Details.} We use AdamW~($\beta_1 \!\! = \!\! 0.95$, $\beta_2 \!\! = 
    \!\! 0.98$)~\cite{adamw} with a base learning rate of $7\cdot10^{-4}$ and a cosine scheduler with linear warm-up. We train for 250\,k iterations with batch size of 32 or the maximum possible. All models are trained with either the standard aggregation-based loss~(CAV)~(\S\ref{sec:vanilla_contrastive}), or the sequence-based one~(SCAV)~(\S\ref{sec:seq_contrastive}), which uses either the Interpolated Euclidean, DTW, or Wasserstein distances. The CAV temperature is initialized to $\tau \! = \!0.07$~\cite{clip}, while for SCAV we found it beneficial to initialize with the value of $\lambda \!= \! 1$. We used a dropout rate of 0.1 and all models were trained on an NVIDIA V100-32GB using full precision.

   \textbf{Evaluation.} We average the 10 best checkpoints according to the validation set performance. We evaluate on the test sets of VGGSound and Music using either aggregation-based, sequence-based, or hybrid retrieval~(\S\ref{sec:retrieval}), and report Audio-to-Visual ($\text{A} \! \rightarrow \! \text{V}$) and Visual-to-Audio ($\text{V} \! \rightarrow  \! \text{A}$) recall@1.
   
    \setlength{\parindent}{12pt}

\section{Results} \label{sec:results}

    \setlength{\parindent}{0pt}

    \textbf{SCAV vs.\ CAV.} In the left part of Table~\ref{tab:preliminary_results}, we present the retrieval results in a setting where CAV and SCAV models are trained with small batch sizes of 32. 
    Initially, we observe that SCAV models perform comparably to CAV when evaluated using standard aggregation-based retrieval. However, their inherent property of being optimized in a sequential distance space allows them to excel in sequence-based retrieval, easily surpassing CAV by up to 10 points in recall. Furthermore, we find that the interpolated Euclidean outperforms DTW and Wasserstein, while also being simpler and more efficient.\footnote{The complexity is $O(n)$ for the Euclidean and $O(n\cdot m)$ for the others.}

    \begin{table}[ht]
        \centering\setlength{\tabcolsep}{2pt}
        \caption{Bidirectional retrieval~(R@1) in VGGSound-Test for CAV and SCAV variants with a small batch size~(BS), and a maxed-out batch size~(max BS). Retrieval is done with the aggregation-based~(Agg) or sequence-based~(Seq) methods. Seq uses the corresponding distance used in SCAV training, while CAV uses the interpolated Euclidean. \underline{Underlined}: best of each column~(Agg or Seq); \textbf{bold}: best overall~(Agg \& Seq)}
        \vspace{-0.1cm}
        \resizebox{\columnwidth}{!}{
        \begin{tabular}{@{}l @{\hspace{0.1cm}} p{-2cm} @{\hspace{0.1cm}} | ccccc @{\hspace{0.1cm}} p{-2cm} @{\hspace{0.1cm}} | ccccc@{}}
        \toprule
        \multirow{2}{*}{\textbf{\begin{tabular}[c]{@{}l@{}}Contrastive\\ Method\end{tabular}}} &  & \multirow{2}{*}{\textbf{BS}} & \multicolumn{2}{c}{$\text{A} \! \rightarrow \! \text{V}$} & \multicolumn{2}{c}{$\text{V} \! \rightarrow \! \text{A}$}                                                                                      &  & \multirow{2}{*}{\textbf{\begin{tabular}[c]{@{}c@{}}max\\ BS\end{tabular}}} & \multicolumn{2}{c}{$\text{A} \! \rightarrow \! \text{V}$} & \multicolumn{2}{c}{$\text{V} \! \rightarrow \! \text{A}$}                                                                                      \\ \cmidrule(lr){4-7} \cmidrule(l){10-13}
                                                                                               &  &                              & \text{Agg}                & \text{Seq}                & \text{Agg}                & \text{Seq}                &  &                                                                            & \text{Agg}                & \text{Seq}                & \text{Agg}                & \text{Seq}                \\ \cmidrule(r){1-1} \cmidrule(lr){3-7} \cmidrule(l){9-13} 
        CAV                                                                                    &  & 32                  & 10.2                        & 10.6                        & \underline{10.3}               & 8.0                         &  & 192                                                               & \underline{12.2}               & 12.7                        & \underline{12.5}               & 11.6                        \\
        + $ \text{pre(v} \! \rightarrow \!\text{a)}$                                         &  & 32                  & 10.0                        & 10.6                        & 9.7                         & 7.9                         &  & 1024                                                              & 12.1                        & 12.2                        & 12.1                        & 11.6                        \\ \cmidrule(r){1-1} \cmidrule(lr){3-7} \cmidrule(l){9-13} 
        $\text{SCAV}_\text{Eucl}$ &  & &  & &  & &  & &  & &  & \\
        
        + $\text{pre(v}\!\rightarrow \!\text{a)}$                              &  & 32                  & 8.7                         & 18.7                        & 7.6                         & 18.3                        &  & 256                                                               & 11.9                        & \textbf{\underline{22.6}}               & 10.1                        & \textbf{\underline{22.3}}               \\
        + $ \text{pre(a} \! \rightarrow \! \text{v)}$                              &  & 32                  & 9.0                         & 18.6                        & 8.1                         & 18.3               &  & 64                                                                & 9.7                         & 20.8                        & 8.3                         & 20.5                        \\
        + $\text{post(v} \! \rightarrow \! \text{a)}$                             &  & 32                  & 9.5                         & \textbf{\underline{19.2}}               & 8.2                         & \textbf{\underline{18.4}}               &  & 96                                                                & 11.4                        & 21.3                        & 7.8                         & 21.3                        \\
        + $\text{post(a}\! \rightarrow \! \text{v)}$                             &  & 32                  & 9.3                         & 18.6                        & 7.7                         & \textbf{\underline{18.4}}               &  & 64                                                                & 10.0                        & 20.6                        & 7.5                         & 20.1                        \\ \cmidrule(r){1-1} \cmidrule(lr){3-7} \cmidrule(l){9-13} 
        $\text{SCAV}_\text{DTW}$                                                               &  & 32                  & 9.2                         & 12.5                        & 9.8                         & 13.0                        &  & 64                                                                & 9.5                         & 14.2                        & 11.5                        & 15.9                        \\
        $\text{SCAV}_\text{Wass}$                                                       &  & 32                  & \underline{11.1}               & 15.0                        & 9.5                         & 14.8                        &  & 64                                                                & 11.3                        & 16.1                        & 9.7                         & 16.6                        \\ \bottomrule
        \end{tabular}
        }
        \label{tab:preliminary_results}
        \vspace{-0.15cm}
    \end{table}
    
    \textbf{Accounting for memory.} To ensure that are comparisons are fair we repeat the previous experiments~(right part of Table~\ref{tab:main_results}) but with the maximum possible batch size for each method.\footnote{A larger batch is generally better due to the presence of more negatives.} Thus we account for the increased memory footprint of some SCAV variants, which stem from the exploding number of point-wise comparisons ($B \! \times \! B \times \! n \! \times \! m$) within a batch. We find that SCAV with Euclidean distance and pre(v$\rightarrow$a) interpolation, which is compressive since audio sequences are generally shorter, outperforms all other methods in both directions. This indicates that having more negative points from additional examples (larger batch size) is more important than more negatives from the current examples (higher sequence resolution). We also experiment with using this compressive interpolation for CAV, allowing it to scale to a batch size of 1024, but due to the absence of intra-sequence modeling, we do not observe any benefits. Due to its simplicity, scalability, and performance, we focus the rest of the experimentation on SCAV with the interpolated Euclidean.
    
    \textbf{SCAV vs.\ Other Methods.} In Table~\ref{tab:main_results} we compare SCAV against two strong baselines from the literature:\footnote{We re-run the provided public models for our test splits. These are the only methods discussed in \S\ref{sec:relevant_works} that have available models.} ImageBind~\cite{imagebind}, which added other modalities to the CLIP space, and CAV-MAE~\cite{cav_mae} which was trained with masked data modeling and aggregation-based contrastive learning using AS2M~\cite{audioset}, a superset of VGGSound. We present results on VGGSound-Test and on the out-of-distribution Music-Test dataset (zero-shot). SCAV variants with interpolated Euclidean outperform the previous methods by large margins, achieving 2$\times$ relative improvements in VGGSound and 3--3.5$\times$ in Music, while utilizing significantly fewer training resources. The larger improvements in Music, which is especially challenging due to the presence of multiple clips with almost identical visual and acoustic features, further supports our assertion regarding the importance of a representation space with proper intra-sequence temporal modeling. Additionally, we observe that different variants are suited for each dataset. We hypothesize that $\text{post(a}\!\rightarrow\!\text{v)}$ is better for Music, because it is expansive, providing more sequential contrastiveness. On the other hand, $\text{pre(v}\rightarrow\text{a)}$ is better for VGGSound, which is compressive, leading to more semantic contrastiveness. Finally, CAV can also outperform the previous methods, which we attribute to the high-quality audio representations of BEATs.

    \begin{table}[t]
    \centering\setlength{\tabcolsep}{2pt}
        \centering
        \caption{Bidirectional retrieval results (R@1) in VGGSound and Music (Zero-shot) test. Upper part of the Table is other works, and lower part is this work. All methods are evaluated on the same test sets. SCAV models use seq-based retrieval.}
        \vspace{-0.1cm}
        \resizebox{0.9\columnwidth}{!}{
            \begin{tabular}{l @{\hspace{0.1cm}} c @{\hspace{0.1cm}} p{-2cm} @{\hspace{0.1cm}} cc  @{\hspace{0.1cm}} p{-2cm} @{\hspace{0.1cm}} cc}
            \toprule
            \multirow{3}{*}{\textbf{Models}}                                     & \multirow{3}{*}{\textbf{\begin{tabular}[c]{@{}c@{}}Train\\ Params\end{tabular}}} &  & \multicolumn{2}{c}{\textbf{VGGSound}}                                                                                 &  & \multicolumn{2}{c}{\textbf{Music}}                                                                                    \\ \cmidrule{4-5} \cmidrule{7-8} 
                                                                                 &                                                                                         &  & $\text{A} \! \rightarrow \! \text{V}$ & $\text{V} \! \rightarrow \! \text{A}$ &  & $\text{A} \! \rightarrow \! \text{V}$ & $\text{V} \! \rightarrow \! \text{A}$ \\ \cmidrule{1-2} \cmidrule{4-5} \cmidrule{7-8} 
            $\text{ImageBind}_\text{HUGE}$                                      & 1\,B                                                                                      &  & 10.8                        & 12.1                        &  & 2.5                         & 2.6                         \\
            $\text{CAV-MAE}_\text{scale+}$                                       & 200\,M                                                                                    &  & 5.5                         & 6.5                         &  & 2.3                         & 2.1                         \\ \cmidrule{1-2} \cmidrule{4-5} \cmidrule{7-8} 
            $\text{CAV}$                                                         & 38\,M                                                                                     &  & 12.2                        & 12.5                        &  & 3.8                         & 2.8                         \\
            $\text{SCAV}_{\text{Eucl/pre(v}\rightarrow\text{a)}}$  & 38\,M                                                                                     &  & \textbf{22.6}               & \textbf{22.3}               &  & 9.3                         & 7.3                         \\
            $\text{SCAV}_{\text{Eucl/post(a}\rightarrow\text{v)}}$ & 38\,M                                                                                     &  & 21.3                        & 21.3                        &  & \textbf{9.5}                & \textbf{8.1}                \\ \bottomrule
            \end{tabular}
        }
        \label{tab:main_results}
        \vspace{-0.3cm}
    \end{table}

    \begin{table}[t]
        \centering
        \caption{Hybrid Retrieval results in VGGSound test with variable pre-selection $k$ and inference time for 1\,k queries in a test set of 10\,k candidates. Results with $\text{SCAV}_{\text{Eucl/pre(v}\rightarrow\text{a)}}$.}
        \vspace{-0.1cm}
        \resizebox{0.78\columnwidth}{!}{
        \begin{tabular}{@{}lccccc@{}}
        \toprule
        \multirow{2}{*}{\textbf{\begin{tabular}[c]{@{}l@{}}Contr.\\ Method\end{tabular}}} & \multirow{2}{*}{\textbf{$k$}} & \multicolumn{2}{c}{\textbf{R@1}}                  & \multicolumn{2}{c}{\textbf{Time}~($\downarrow$)} \\ \cmidrule(l){3-6} 
                                                                                          &                               & $\text{A} \! \rightarrow \! \text{V}$ & $\text{V} \! \rightarrow \! \text{A}$ & GPU             & CPU             \\ \midrule
        \multirow{5}{*}{SCAV}                                                             & 10$^{4\dagger}$              & 22.6                    & 22.3                    & 13.9            & 741            \\
                                                                                          & 10$^{3\phantom{*}}$               & 22.6                    & 22.3                    & 1.7             & 7.0             \\
                                                                                          & 10$^{2\phantom{*}}$                  & 22.6                    & 22.1                    & 0.5             & 4.6             \\
                                                                                          & 10$^{1\phantom{*}}$                  & 20.8                    & 19.1                    & 0.4             & 3.6             \\
                                                                                          & 10$^{0*}$               & 11.9                    & 10.1                    & 0.3             & 2.6             \\ \midrule
        CAV                                                                               & -                             & 12.2                    & 12.5                    & 0.3             & 2.6             \\ \bottomrule
        \multicolumn{6}{l}{$^\dagger$ Sequence-based retrieval. $^*$ Aggregation-based retrieval.}
        \end{tabular}
        }
        \label{tab:hybrid}
        \vspace{-0.3cm}
    \end{table}

    \textbf{Hybrid Retrieval.} In Table~\ref{tab:hybrid} we evaluate the performance of the proposed hybrid retrieval (\S\ref{sec:retrieval}), by varying the number of pre-selected candidates $k$. Our results on VGGSound show that SCAV with hybrid retrieval easily surpasses CAV, and reaches the performance of sequence-based retrieval with as few as $k\!=\!10^2$ candidates. Furthermore, the inference time measurements demonstrate the efficiency of the hybrid method. The retrieval process is accelerated drastically, from the 13.9 GPU seconds required with sequence-based, down to 0.5 seconds for $k\!=\!10^2$, a negligible computational overhead compared to aggregation-based retrieval~(CAV, 0.3 seconds).

    \textbf{Ablations.} In the ablation (1) of Table~\ref{tab:ablations}, we find that normalizing the distance matrix (\S\ref{sec:seq_contrastive}) of $\mathcal{L}^\text{seq}$ is essential, since without it, the sequential distances are not very distinct from each other, and performance is on par with CAV. Next in (2), we show that for SCAV is better to start training with a larger temperature of $\lambda \! = \! 1$, as opposed to the usual 0.07~\cite{clip}, which makes the task initially harder, but leads to better performance. Finally, no gains can be observed by combining both the CAV and SCAV objectives, indicating that SCAV alone is sufficient for both temporal and semantic modeling.

    \begin{table}[t]
        \caption{Ablations. Bidirectional retrieval in VGGSound-Test.}
        \vspace{-0.1cm}
        \centering\setlength{\tabcolsep}{2pt}
        \begin{tabular}{@{}lcc@{}}
        \toprule
        \textbf{Method}                                      & $\text{A} \! \rightarrow \! \text{V}$  & $\text{V} \! \rightarrow \! \text{A}$ \\ \midrule
        $\text{SCAV}_{\text{Eucl/pre(v}\rightarrow\text{a)}}$ & \textbf{22.6}     & \textbf{22.3}     \\
        $\hookrightarrow$ (1) w/o Distance Matrix Norm.                   & 12.5              & 12.7              \\
        $\hookrightarrow$ (2) w/ $\lambda \! = \! 0.07$ Init.        & 18.6              & 18.5  \\
        $\hookrightarrow$ (3) w/ CAV \& SCAV multi-tasking          & 15.3              & 15.4 
        \\ \bottomrule
        \end{tabular}
        \label{tab:ablations}
        \vspace{-0.35cm}
    \end{table}

\section{Conclusion}

    We presented a novel approach for audio-visual contrastive learning based on sequential distances, which also enables their use during inference for retrieval. We investigated several distance metrics, and showed that our method with an interpolated Euclidean distance surpasses aggregation-based contrastive learning and other methods by large margins. Furthermore, we demonstrated that our method is particularly effective in a challenging out-of-distribution setting where enhanced intra-sequential discrimination capabilities are needed. Finally, we proposed a hybrid method for retrieval that is both efficient and powerful, and opens the way for our method to be performant in large-scale retrieval scenarios. Future research will explore the addition of the text modality in the sequential audio-visual representation space, which can enable many multimodal applications, like text-guided video generation.

\vfill\pagebreak

\let\oldbibitem\bibitem
\renewcommand{\bibitem}[1]{\oldbibitem{#1}\setlength{\itemsep}{0pt}}

\bibliographystyle{IEEEbib}
\bibliography{references}

\end{document}